\documentclass[conference]{IEEEtran}
\IEEEoverridecommandlockouts
\usepackage{cite}
\usepackage{algorithmic}
\usepackage{graphicx}
\usepackage{textcomp}
\usepackage[dvipsnames]{xcolor}
\usepackage{booktabs} 
\usepackage[utf8]{inputenc}
\usepackage{textcomp}
\usepackage{balance}
\usepackage[ruled, lined, linesnumbered, commentsnumbered, noend]{algorithm2e}
\usepackage{amsmath,amssymb,amsfonts,amsthm}
\usepackage{thmtools} 
\usepackage{thm-restate}
\usepackage{enumitem}
\usepackage{multirow}
\usepackage{pifont}
\usepackage{color}
\usepackage{colortbl}
\usepackage{multicol}
\usepackage{booktabs, tabularx, makecell, multirow,caption}
\usepackage{xspace}
\usepackage{graphicx}

\newcommand{\system}{\emph{DecentPeeR}\xspace}
\newtheorem{myTheorem}{Theorem}
\newtheorem{myLemma}{Lemma}

\newtheorem{myDefinition}{Definition}
\newtheorem{myAssumption}{Assumption}
\newcommand{\xmark}{\ding{55}}%

\def\BibTeX{{\rm B\kern-.05em{\sc i\kern-.025em b}\kern-.08em
    T\kern-.1667em\lower.7ex\hbox{E}\kern-.125emX}}
\begin{document}

\title{DecentPeeR: A Self-Incentivised \& Inclusive \\ Decentralized Peer Review System}

\author{\IEEEauthorblockN{Johannes Gruendler$^1$ \quad Darya Melnyk$^1$ \quad Arash Pourdamghani$^{1,2}$ \quad Stefan Schmid$^{1,2}$}
\IEEEauthorblockA{$^1$TU Berlin, Germany $^2$Weizenbaum Institute, Germany}
}
\maketitle

\begin{abstract}
Peer review, as a widely used practice to ensure the quality and integrity of publications, lacks a well-defined and common mechanism to self-incentivize virtuous behavior \emph{across} all the conferences and journals. This is because information about reviewer efforts and author feedback typically remains local to a single venue, while the same group of authors and reviewers participate in the publication process across many venues. Previous attempts to incentivize the reviewing process assume that the quality of reviews and papers authored correlate for the same person, or they assume that the reviewers can receive physical rewards for their work.
In this paper, we aim to keep track of reviewing and authoring efforts by users (who review \emph{and} author) across different venues while ensuring self-incentivization.

To this end, we introduce \system, a system that captures the interactions of users who use a peer review system as decentralized reputation scores.
We show that our system incentivizes reviewers to behave according to the rules, i.e., it has a unique Nash equilibrium in which virtuous behavior is rewarded.
Furthermore, we detail how our design ensures inclusivity, i.e., giving everyone a fair chance to publish, especially when facing 
dishonest
users. We also report on empirical results that show the incentive mechanism works:   dishonest individual and group behavior are penalized, but it is possible to recover from a poor score over time.
\end{abstract}


\section{Introduction}
Peer review systems are widely used and have an extensive impact in today's academia.
Use cases of peer-review ranges from the scientific publication process~\cite{ross2017survey,walker2015emerging} to open source software development~\cite{Software,Software12}. 
With the popularity of peer review systems, various methods have been proposed to make the peer review procedure more inclusive. With inclusivity, authors have a chance to publish their work solely based on its quality. Ensuring inclusivity is challenging in the academic peer review process: submissions on different research topics may not be comparable; reviewers may have personal opinions depending on the topic of the submission; due to large amounts of published papers, evaluations from only few reviewers can be used decide on the quality of a submission.
Previous efforts to ensure inclusivity range from enforcing prior-announcement of conflict-of-interest~\cite{radun2021nonfinancial,ancker2007comparison}, double-blindness~\cite{tomkins2017reviewer,DoubleBlind2}, and actions from the editor to promote quality, integrity, and fairness~\cite{ResnikE16}. 
Most traditional solutions are focused on how to make a \emph{single} conference more inclusive. However, authors and reviewers likely take part at multiple venues over their careers. Thus, a cross-venue measure would be more viable today, given the advancements in decentralized technologies.
 
In this work, we keep track of the actions of users \emph{over time} using a \emph{reputation system} to ensure inclusivity. We build a decentralized system where users tend to follow the rules of the system based on their best interests. While incentivizing users who behave rationally, the system should not punish academic work of good quality and thus violate inclusivity.
To this end, we develop a self-incentivized system based on a game theoretical approach, showing that achieving the unique  Nash equilibrium is only possible by adhering to the rules of the system.
We detail our system, \system, from the perspective of an academic who wants to contribute to or organize a conference. 
In doing so, we also benefit from the decentralized storage mechanisms provided by blockchain technology~\cite{IPFS19,tron2020book} 
that is used
to keep a history of peer review systems' data across different venues.

We detail our system, \system, from the perspective of an academic who wants to contribute to or organize a conference. 
In doing so, we also benefit from the decentralized mechanisms provided by blockchain technology (e.g., to keep a history of peer review systems' data across different venues).

\subsection{Design Goals}

Our aim is to design our peer-review system that satisfies the following main goals:

\noindent \textbf{Self-incentivization.} Typically, peer review systems assume that participants are well-behaved~\cite{7424543}, that reviewers can be assigned a reliability score~\cite{NIPS2011_c667d53a}, or that at most a small percentage of reviewers is biased~\cite{10.1145/2505057,Sadler2006TheIO}. However, these rules might be neglected, resulting to the raise of adversarial reviews~\cite{AdvReview}.
As a remedy, we aim to ensure that it is in the reviewer's own best interest to respect rules of the system, i.e., through a provable self-incentivization.

\noindent \textbf{Inclusivity.} 
A reputation system should be flexible and adaptive and in particular also support new users of the system. 
Our goal is to design a system where everyone would have a chance to contribute good quality work, mostly independent of their scores, and where users with bad scores have a chance to recover from a failure. 

\noindent \textbf{Cross-venue evaluation.} 
Academic users do not encounter peer-review systems only once: they participate in multiple venues (conferences, journals, workshops, etc.) throughout several years. Hence we aim to design a system that benefits from this fact and also aggregates data over time.

\subsection{Contribution}

In this paper, we design a self-incentivized and inclusive peer review system, \system, that works across venues and tracks.
Honest reviewing behavior is rewarded
with a positive influence on future borderline-scored submissions by the authors. In addition, the reviewing score provides committee chairs with a criterion to select the committee.

We present a peer review game in this paper and show that it has a unique Nash equilibrium where the 
users play honestly.  We then analyze the desired properties of the peer review game, such as inclusivity and cross-venue evaluation.
Our method incorporates three mechanisms that lead to a high level of inclusivity:
\begin{itemize}
    \item Our system only considers the reputation score in borderline cases: if the high quality of a paper is already agreed upon, we consider that paper as accepted. 
    \item Our system uses a randomness mechanism to form a program committee and a reviewing team: the weighted randomness ensures that selected users can be trusted while giving chance to every user to participate. 
    \item A reputation score function has been implemented with the goal of ensuring fast recovery for users who have limited misbehavior. 
\end{itemize} 

We show that the cross-venue aspect allows us to quickly detect adversarial behavior by reviewers. Once an adversarial reviewer however behaves correctly again, the corresponding score of the reviewer recovers. We conclude by showing that a majority attack, where adversarial reviewers collaborate in order to evaluate the paper dishonestly, is unlikely under the uniformly chosen reviewers in the review assignment. 


Our paper is organized as follows: we follow by relating our paper to previous works. We then detail \system design in \S \ref{sec: design} and analyze it in \S \ref{sec: analysis}. Finally, we go over a few case studies in \S \ref{sec: case} and conclude our work in \S \ref{sec: conclude}. 

\subsection{Related Work}
\label{sec: related work}

Peer review is a broad research topic that has been investigated from many aspects over the past years. For example, empirical studies have been conducted on peer review for classroom use~\cite{ChanTsenChou20113i,e567bb29-7acf-3456-a64a-627506a1cc95}, for conference reviews~\cite{shah2018design, cortes2021inconsistency}, and for funding applications~\cite{vallee2022peer}.   
In the following, we discuss different perspectives on peer review systems, coming from theoretical research
as well as practical systems.

\noindent \textbf{Blockchain-based peer review.}
Our 
peer review system 
can be implemented on a blockchain-based system that supports smart contracts, like Ethereum. The decentralized nature of blockchain-based protocols has previously motivated many researchers to utilize its advantages for new designs of peer review systems. Our work is also a building block towards a~Decentralized Science (DeSci) future~\cite{DingHLGQKW22}.

Initial proposals for an alternative blockchain-based peer review systems ~\cite{SharplesD16, PubChain} focus on providing an alternative coin instead of Bitcoin. These attempts to alter a financial system for peer review, however, are inherently flawed: wealthy participants can game the system to their benefit.

Another set of blockchain-based peer review systems focused on providing a decentralized platform to store information exchanged during a peer review process~\cite{NizamuddinHS18, IPFS19}, mostly leveraging the advantages provided by IPFS. Although it is possible to use the proposed peer review systems across venues, the systems lack the essential requirements that ensure fair treatment of all users when being used across venues. 

More recent systems~\cite{ICBC21Review,ICBC22} aimed to tackle the challenge of a collaborative system. In doing so, they showed what are possible ways to provide self-invitation based on a game-theoretic perspective. In a nutshell, they showed that the allowed rules of the game are in the best interest of all users. However, they did not show what happens when a failure happens and how the system could recover. Furthermore, they did not consider the fact that not only a single venue exists, and a system should not be restarted whenever a new request for a venue appears. We summarize the main properties of the presented distributed peer review systems in Table~\ref{tbl: overview}.

\begin{table}[t]
\begin{center}
\begin{tabular}[t]{p{0.195\textwidth}|>{\centering}m{0.038\textwidth}|>{\centering}m{0.084\textwidth}|c}
     System &
     Cross-venue & Self-incentivising & Inclusive \\\hline
PubChain~\cite{PubChain} &\cellcolor{green!20} \checkmark & \cellcolor{red!20} \xmark& \cellcolor{red!20} \xmark\\
\hline
Blockchain and Kudos~\cite{SharplesD16} & \cellcolor{green!20} \checkmark  & \cellcolor{red!20} \xmark   &  \cellcolor{red!20} \xmark\\
\hline
IPFS-based ~\cite{NizamuddinHS18} & \cellcolor{green!20} \checkmark  & \cellcolor{red!20} \xmark   &  \cellcolor{red!20} \xmark\\
\hline
Data Marketplaces~\cite{ICBC21Review}  &\cellcolor{red!20} \xmark& \cellcolor{green!20} \checkmark & \cellcolor{red!20} \xmark\\
\hline
\cellcolor{white!20} Collaborative Research~\cite{ICBC22}  & \cellcolor{red!20} \xmark& \cellcolor{green!20} \checkmark& \cellcolor{red!20} \xmark
\\
\hline
\cellcolor{gray!20}  \system (our work) & \cellcolor{green!20} \checkmark & \cellcolor{green!20} \checkmark & \cellcolor{green!20} \checkmark
\end{tabular}
\caption{Comparing \system with previous decentralized peer review systems.}
\label{tbl: overview}
\end{center}
\end{table}

\noindent \textbf{Social choice perspective on peer review.}
In social choice studies, peer review has been investigated as an assessment method for 
grading homework and exams\cite{Stelmakh_Shah_Singh_2021}, programming classes\cite{10.1145/1531674.1531692}, and conferences\cite{2dd8e457791c474fab36c6a9edd09054}.

Some of these works focus on finding the right aggregation rules, by investigating cardinal voting rules instead of ordinal grading~\cite{10.1145/3412347,10.1145/2623330.2623654} or by showing specific properties of peer review protocols, such as strategyproofness~\cite{2dd8e457791c474fab36c6a9edd09054}. Other papers focus on determining truthful reviewers by assuming that the grading is performed with assistance~\cite{10.1145/2676723.2677278}.  There are also studies on reputation-based peer grading. Leaning on PageRank, Walsh proposed a PeerRank system~\cite{Walsh2014ThePM}. In this system, each reviewer gets a reputation score, which is computed by the grades given by the reviewer weighted by the reputation score of the reviewer. Note that a reputation score calculation is connected to the grade of the submitted work, which is not necessarily true in conference reviewing.

\noindent \textbf{Reputation systems.}
Reputation systems are one of the key tools to establish trust in an untrustworthy environment~\cite{VavilisPZ14}.
For example, they have an established place in designing distributed systems, especially in connecting peers in a peer-to-peer network~\cite{HendrikxBC15}. They have also been considered in other applications such as e-commerce~\cite{KugblenuVK21} and transportation~\cite{YangZ0L17}.
Most of the previous work that considers a reputation system on blockchain~\cite{RepuCoin19JiangshanY,ProofOfReputation} uses the reputation score as an alternative mining protocol, which is orthogonal to how we used it in our system.

\noindent \textbf{Similarity detection mechanisms.}
The digital age has made it easier to reuse the efforts of others, and hence, from early on, it was critical to create similarity detection mechanisms~\cite{SimilarityBrinDG95}. Also, for the peer review process similarity detection is inevitable, as it enables editors to verify the integrity of a research work.  With the rise of generative AI tools, advanced similarity detection mechanisms received even more attention, especially because it became possible to rehash ideas of others into undetectable results (even by trained eyes)~\cite{SimilarityCrothersJV23}. 
To detect such behavior, context-oblivious methods have been proposed, such as fingerprinting~\cite{SimilarityFingerPrintSchleimerWA03}, text matching~\cite{SimilarityKarnalimSC19}, or compression~\cite{SimilarityFrankCW00}. The latter one can run fast but might have low accuracy due to false negatives.
In general, context-aware similarity detection approaches allow us to reduce false negatives, but they require much more computational power. As an alternative, latent semantic analysis~\cite{SimilarityCosmaJ12} or transformer-based tools~\cite{SimilarityGiovannottiG21} have been considered in the literature. In this work, we do not focus on a particular similarity detection mechanism, but rather assume that such methods exist and can be deployed by the system designer.

\section{System Design}
\label{sec: design}
This section presents the design of \system.
Our system consists of a set of \emph{users} that have two roles: \emph{authoring} and \emph{reviewing} papers.
We use $N$ to denote the set of all $n$ users in the system. Every user in this mechanism has a \emph{reputation score} that is initially set to $R_i^t = \frac{1}{2}$ for a user $i$ and time $t_i$. As we later show, the reputation score can influence the status of the paper and the users themselves.
For any given paper, multiple authors might write a paper or multiple reviewers check a paper. To have a simplified notation, we introduce the \emph{unified} reviewer and author.

\begin{myDefinition}[Unified Author \& Unified  Reviewer]
\label{def: Total Reviewer} 
A \emph{unified author} is the set of authors who are submitting their work together. Similarly, a  \emph{unified reviewer} is the set of reviewers who check a paper.
\end{myDefinition}

We design our system to be inclusive, i.e., encourage honest behavior (regardless of users) while allowing a chance to recover from faulty behaviors.

\begin{myDefinition}[Honest Behavior \& Faulty Behavior]
\label{Def2 : Faulty Behaviour} 
Honest behavior denotes an outcome that respects the rules of the venue set by the program committee. The opposite of honest behavior we call faulty behavior.
\end{myDefinition}

\subsection{System Setup}
We now describe the setup of \system, which is designed with the real-world implementation in mind. 

\subsection{Blockchain-Based Implementation}
In the following we detail how our system benefits from the decentralized structure of blockchain systems. In particular, how we keep track of the venues' data and allow users to interact with the data.

\noindent \textbf{Storage on blockchain.}
We benefit from the blockchain storage capabilities in two ways: Firstly, by storing the papers' and the users' metadata (e.g., name) on a public ledger, so they can be retrieved and used easily in the future. Secondly, we keep the confidential information about papers (e.g., their content) and users (e.g., their reputation) encrypted in a second-layer storage system like IPFS~\cite{IPFSSurvey22}.

\noindent \textbf{Organization via smart contracts.}
Our peer review system benefits from
smart contracts~\cite{kemmoe2020recent,hewa2021survey} that are used to implement its core functionalities.
Actions such as registration, random generation for reviewer assignment (using tools such as ~\cite{Prob-Smart-contract19,qian2023demystifying}), and paper submissions are handled via the smart contract. It plays a vital role in safely decentralized tracking of the behavior of the users, reputation scores, and topic tags. In addition, it allows to implement the system across venues. The 
possible required payments
for maintaining the contract is considred to be included in the conference fee.

\subsection{Conference Processes}
\begin{figure}[tb]
    \centering
    \includegraphics[width=0.9\linewidth, clip]{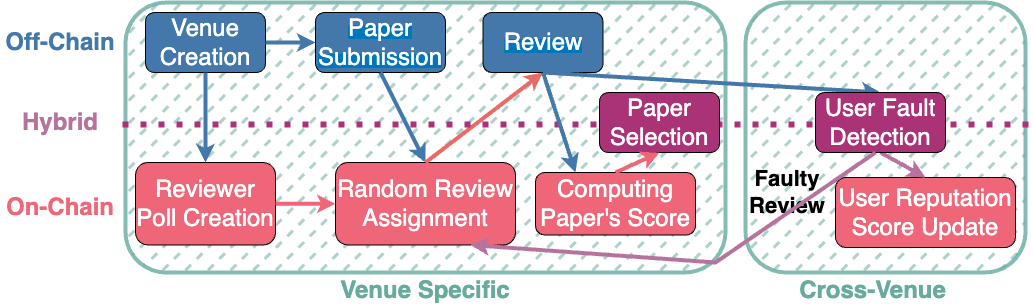}
    \caption{ \system processes and their division into an off/on-chain as well as venue-specific/cross-venue. 
    }
    \label{fig: Lifetime}
\end{figure}

These processes are detailed below, overviewed in 
Figure~\ref{fig: Lifetime}.  

\noindent \textbf{Venue \& reviewer pool creation.} To initiate a new venue (or add a new iteration of the venue), program chairs can create a venue-specific instance. Based on the topics that they mentioned in the definition of the conference, the system suggests a pool of reviewers with high reputation who have indicated similar topics to the conference description.

\noindent \textbf{Paper submission \& reviewer assignment.}
When a unified author wants to submit a paper, we consider the case that a part of their reputation is stored as a deposit to avoid spam submissions. 
After the successful submission of a paper, a unified reviewer is chosen uniformly at random from the set of reviewers (excluding the authors and conflicts of interest) for this paper. Then reviewers get a call to review individually. If they accept this call, they should complete the review in the given time frame.

To balance the unified reviewer based on their expertise in the respective area, a \emph{confidence score} is calculated for each reviewer. Such a score can be calculated with any of the similarity detection techniques mentioned previously.
One can compare the degree of similarity between the tags provided by a reviewer with the tags of the paper that need to be reviewed. Let us consider the output of such a similarity detection as $\sigma(T_R,T_P)\in[0,1]$, where $T_R$ and $T_P$ are the reviewer's and the paper's tags. Hence, the total competence of a reviewer $j$ reviewing paper $p$ is $C^p_j:=\sigma(T_R,T_P)$.

\noindent \textbf{User fault detection.}
If users do not adhere to the rules, this behavior should be detected by a peer review system. One example is the submission of identical reviews for a new version of the paper. In our cross-venue system, comparison to older review versions becomes possible, for example, by using similarity detection mechanisms discussed before.

\noindent \textbf{Weighted score of a paper.}
Consider a paper which is scored by $r$ reviewers with respective scores $S^p_j \in [1,5]$ for reviewers $j\in\{1,..,r\}$. We say that a paper is honest if the average score is above a preset threshold.

A borderline paper 
(that has not been clearly accepted or rejected by the reviewers)
receives a weighted score $W^p$. This score is based on the reputation of the unified author $R_i^t$ and the score for the paper $S^p_j$ by reviewer $j$. The weighted score is calculated as follows:
\[
W^p := R_i^t\cdot\frac{\sum_{j=1}^r C^p_j \cdot R_j \cdot S^p_j }{r \cdot \sum_{j=1}^r C^p_j\cdot R_j}
\]

The link between the score of a paper and the reputation of a user is an incentive to gain a high reputation score.

\noindent \textbf{Paper selection.} Papers that receive a score above a predefined threshold for acceptance are accepted directly, and similarly for papers with very low score. 
Besides these two extremes, there might exist a set of borderline papers. 
If there is still a possibility to accept more papers, our method chooses those based on the average reputation score of authors. 
This motivates the reviewers to submit high-quality reviews, as a high reputation leads to acceptance of their own borderline-scored papers.
Furthermore, it is still possible for all authors to be part of the scientific community if their papers are good enough, despite their poor review reputation. 

\noindent \textbf{Reputation score update.}
A user's reputation score is updated at the end of the reviewing process of any given venue. The honest behavior of a user leads to a higher score, and a faulty behavior leads to a score reduction. The details on how the reputation score is defined and updated is provided next.

\subsection{Details of Reputation Score Update}
\label{sec : detail repu score}
The reputation score of user $i$ is updated at the start of a new time interval $t$, and only updated if the user $i$ is \emph{active}, i.e., had at least one review or paper in the previous time interval.

The \emph{punishment factor} $P^t_i$ will decrease the reputation of $i$, according to number of times and against whom user $i$ showed faulty behavior. A unified author behaves faulty towards a unified reviewer and vice versa. Based on the current reputation, a \emph{gain summand} $G^t_i$ is added. The reputation score of user $i$ for the next time interval $t+1$ is calculated as:
\begin{equation}
\label{form: main}
     R_i^{t+1} := R_i^t \cdot P^t_i + G^t_i
\end{equation}

Above definition is an overview of the reputation update, which will be detailed next (a summary of the notation used in our paper is provided in Table~\ref{table: parameters} for clarity). 
Let $D_i$ be the \emph{interaction set}, which contains all the users that user $i$ interacted with in the current time interval. The set $D_i^H$ is the subset of users that $i$ interacted honestly with, and $D_i^F$ is the set of users that $i$ behaved faulty against.
At first, the punishment factor is specified. It punishes faulty behavior in reviewing and proposing.
\[    
    P^t_i := \frac{1 + \sum_{k\in D_i^H} R_k^t}{1 + \sum_{k\in D_i} R_k^t}
\]
If user $i$ behaves faulty against a user $j$ with a high reputation, the punishment increases because $R_j^t$ appears only in the 
denominator. High-reputation users provide honest contributions 
over time to the mechanism, and wasting their effort is punished 
harder. Furthermore, their evaluation is considered of higher quality and is therefore allowed to have a greater impact. The unified reviewer acts like one reviewer that an 
author interacted with. The same is true for a reviewer interacting 
with a unified author. Thus, only the reputation of the total 
reviewer or unified author is added and not the reputation of every individual reviewer. If user $i$ was honest in all interactions 
in the last time interval, the punishment score is set to $1$, and no 
punishment takes place. Otherwise, $P^t_i<1$. Therefore,  
one is guaranteed to only be punished by $P^t_i$ in case of at least one faulty action. 

\begin{table}[t]
\begin{center}
\begin{tabular}{ c | c | c}
   \hline 
   Var. & 
   \cellcolor{gray!20} Meaning
    & Domain
     \\ \hline
    \cellcolor{gray!20} $t_i$ & The number of time intervals & $(0,\infty)$
    \\ \hline
    \cellcolor{gray!20} $R_i^t$ &
    Reputation Score of the user&
    $(0,1)$
      \\ \hline
         \cellcolor{gray!20} $P^{t,X}_i$ &
   Punishment factor for behavior $X$ & (0,1]
   \\ \hline 
     \cellcolor{gray!20} $G^t_i$ & Gain in reputation& $(0,\alpha]$
   \\ \hline
    \cellcolor{gray!20} $S^p_j$ &Score of a paper $p$ by reviewer $j$ & $[1,5]$
    \\ \hline
    \cellcolor{gray!20} $W^p$ & Weighted average score for a paper $p$ & $[0,5]$  
    \\ \hline
    \cellcolor{gray!20} $C^p_j$& Competence to of reviewer $j$ to review paper $p$ & [0,1]
    \\ \hline
\end{tabular}
\end{center}
\caption{Variables 
used in our mechanism, their meaning and domain. Index $i$ indicates a user, and $t$ the time.
}
\label{table: parameters}
\end{table}

Next, we discuss the gain received by the users. The gain summand $G^t_i$ rewards a user based on the reputation score $R_i^t$ at time $t$, the punishment factor $P^t_i$, and the time $t_i$ that user $i$ spent active. To this end, the function $f$ is defined as:
\begin{align*}
\label{Def: function f}
    &f(x,t_i) :=
    \begin{cases} 
        \alpha(2x)^{3-g(t_i)}, & x\in \bigl(0,\frac{1}{2}\bigr] \\
        \alpha(2(1-x))^{3-g(t_i)}   , & x\in \bigl(\frac{1}{2},1\bigr)
    \end{cases}
\end{align*}
We further define $G^t_i$ as
$G^t_i := f(R_i^t\cdot P^t_i,t_i)$.

The function $g(t_i) := \sum_{k<t_i} \frac{1}{2^{k}}$ shifts the exponent of $f$. If $t_i$ grows, $g(t_i)$ asymptotically approaches $2$ from below and is based on the geometric series. The 
reward at the domain boundaries of the reputation score is increasing over time. This rewards users
who contributed more work to the system, since $t_i$ only increases with each contribution.

Moreover, \system is designed to be welcoming. Reviewers are drawn randomly from the set of users in the system and the reviewers' expertise may not be on the level of an expert in the field. Hence, the expertise of a reviewer to analyze a paper is given by the comparison of the tags of the paper that is reviewed and the tags of the papers a reviewer published earlier. We use a sequential recommendation transformer to reshape the data of the tags into vectors. We compare the degree of similarity between the tags of the reviewer $T_S$ with the tags of the paper $T_P$ that has to be reviewed. The output of cosine similarity detection is bounded by $\sigma(T_S,T_P)\in[0,1]$. With this measurement, we adjust the punishment based on the knowledge of the reviewer of the field.
\[
\Tilde{P}_i := P^t_i+(1-P^t_i)\left(1-\sigma(T_S,T_P)\right)
\]

\subsection{Peer Review Game}
\system has a \textit{normal form game}
in its heart. A normal form game
consists of a finite set of users $I$, where each user $i$ has a 
finite set of pure strategies $S_i=\{1,...,n\}$ to choose from. The 
collection of pure strategies for all users $s=(s_1,s_2,...,s_n)$ is 
called a \textit{strategy profile}. Every user $i$ has to choose one pure 
strategy $s_i \in S$. The payoff $u_i(s)$ for $i$ is given by the entry corresponding to
$s$ in the utility matrix or tensor of user $i$. 

\begin{myDefinition}[Nash equilibrium]
We say $s$ is a Nash equilibrium if for any user $i \in I$ there exists no 
$s^\prime_i \neq s_i$ such that $u_i(s^\prime_i,s_{-i})>u_i(s)$. Here $s_i$ is the strategy of player $i$ and $s_{-i}$ is the strategy of other players.
\end{myDefinition}
In other words, in a Nash equilibrium, no user can increase their own utility by playing another pure 
strategy, while all other users do not change their strategy. 
Nash equilibria are stable states of a game, where no user has an incentive to 
deviate from their own current strategy.

In \system, users  
can either act honestly or faulty. Let $H$ denote the honest and $F$ denote the faulty behavior, where user $i$ submits a paper and user $j$ reviews. The modeling as a game can be seen in Table~\ref{table: Peer Review Game}.
The outcomes $X,Y$ for $i$ and $A,B$ for $j$ denote the new reputation score according to the corresponding strategy. To differentiate these cases, we denote $P^{t,H}_i$ or $P^{t,F}_i$ to be the punishment factor for the strategy chosen by $i$. We define $X$ to be the outcome of an honest user $i$ and $A$ the honest outcome of user $j$. If a user is faulty, the outcome is defined to be $Y$ for user $i$ and $B$ for $j$. To analyze the outcome, we consider the set $D$ denoting the interactions in the last time interval. We assume that the interaction between $i$ and $j$ is not contained in $D$ and consider the outcome for the respective behavior:
\[
P^{t,H}_i = \frac{1 + \sum_{k\in D_i^H} R_k + R_j}{1 +\sum_{k\in D_i} R_k +R_j};
\ 
P^{t,F}_i = \frac{1 + \sum_{k\in D_i^H} R_k}{1 +\sum_{k\in D_i} R_k +R_j}
\]
For user $i$ the outcome is defined as follows:
\[
X := R_i^tP_i^H + f(R_i^tP_i^H,t_i);
\quad
Y := R_i^t P^{t,F}_i + f(R_i^t P^{t,F}_i,t_i)
\]
For user $j$ it is defined as
\[
A := R_j^tP_j^H + f(R_j^tP_j^H,t_j);
\quad
B := R_j^t P_j^F + f(R_j^t P_j^F,t_j)
\]

\section{Analysis}
\label{sec: analysis}
In this section, we detail how \system ensures the necessary requirements of a peer review we laid out, in particular, we focus on self-incentivization and inclusiveness.
Before discussing the details, we state the assumptions that should be satisfied by any scientific venue. We first assume that any user is motivated to eventually publish at some venue.
\begin{myAssumption}
~\label{assumption: propose} 
Among an infinite number of submissions, there exists an infinite number of submissions from any user.

\end{myAssumption}
Based on this assumption, we considered a user to be both an author and a reviewer, eventually.
Next, we formalize our discussion about the existence of a context-based checking tool for plagiarism or other academic misconduct. We thereby assume that the fault detection mechanisms are not perfect; with probability $\pi$ the detection mechanism might fail.
We would like to emphasize that we only considered the existence of such a misconduct detector oracle, a suitable mechanism can be selected based on the conference needs.

\begin{table}[t]
    \setlength{\extrarowheight}{2pt}
    \begin{center}
    \begin{tabular}{cc|c|c|}
      & \multicolumn{1}{c}{} & \multicolumn{2}{c}{User $j$}\\
      & \multicolumn{1}{c}{} & \multicolumn{1}{c}{$H$}  & \multicolumn{1}{c}{$F$} \\\cline{3-4}
      \multirow{2}*{User $i$}  & $H$ & $(X ,A)$ & $(X,B)$ \\\cline{3-4}
      & $F$ & $(Y,A)$ & $(Y,B)$ \\\cline{3-4}
    \end{tabular}       
    \end{center}
    \caption{A general structure of the game used in this paper.}
    \label{table: Peer Review Game}
\end{table}

\begin{myAssumption}
\label{assumption: cheating detection} 
There exists a mechanism to detect faulty behavior, with probability of failure $\pi$. 
\end{myAssumption}
Given the assumptions above, we now go over the proof that shows \system adheres to peer review requirements.

\subsection{Self-incentivization}
\label{sec: incentiv}
In order to prove that our system has the self-incentivization property, we show that playing honestly for both the author and reviewer is the best possible strategy. 
In other words, we show that the outcome $(H,H)$ (where the author and the reviewer are playing honestly) is a pure Nash equilibrium. 

\begin{myTheorem}
\label{Theorem 1}
Let $\alpha<\frac{1}{6}$.
The outcome $(H,H)$ is the unique pure Nash equilibrium of the peer review game.
\end{myTheorem} 

In order to prove Theorem~\ref{Theorem 1}, we need to show that the reputation score always goes down after a faulty behavior (and vice versa). This must be true for any reputation score and punishment factor.

\begin{myLemma}
\label{lem: honest-play}
Let $x,y \in (0,1)$, $x<y$ and $\alpha<\frac{1}{6}$.
Then the following inequality holds:
\begin{align}
\label{lem 1 statement}
    f(x,t_i) + x > f(y,t_i) + y
\end{align}
\end{myLemma}
\begin{proof}
In the rest of this proof, we fix a $t_i$ and remove it from the notation for clarity. Furthermore, the exponent in $f$, i.e. $3 - g(t_i)$, is substituted by $b\in (1,3]$ in order to simplify $f$. Hence, the Lemma~\ref{lem 1 statement} can be written as 
\[
    h(x) := f(x) + x > f(y) + y := h(y). 
\]
Proving the lemma reduces to showing $h(x)>h(y)$. As $x$ is always bigger than $y$ (given $R_i^t<P^t_i$), we only need to show that $h^\prime(z)>0$ for $z\in[0,1]$.

We next compute the derivative of $h(z)$, $h'(z)$:
 \[
 h^\prime(z) =
 \begin{cases} 
    \alpha2^b b z^{b-1} +1, & z\in \bigl[0,\frac{1}{2}\bigr] \\
    -\alpha2^b b(1-z)^{b-1} + 1   , & z\in \bigl(\frac{1}{2},1\bigr]
 \end{cases}
 \]

For $z\leq \frac{1}{2}$, we can see that $h^\prime(z) > 0$ holds since all variables and parameters are positive and 1 is added. 

In case of $z>\frac{1}{2}$, given $b\in(1,3]$ and $\alpha<\frac{1}{6}$, we have: 

\begin{align*}
 \alpha<\frac{1}{2b} \Rightarrow(\alpha2^b b) < 2^{b-1}
 \Rightarrow  \frac{1}{2} > 1-\frac{1}{(\alpha2^b b)^\frac{1}{b-1}}
\end{align*}
As we know that $z > \frac{1}{2}$ therefore we have
 \begin{alignat*}{2}
 &z > 1 - \frac{1}{(\alpha2^b b)^\frac{1}{b-1}}
 \Rightarrow
 (1-z)^{b-1}<\frac{1}{\alpha2^b b}  \\
 &\Rightarrow \alpha2^b b(1-z)^{b-1} < 1 
 \end{alignat*}

and hence $h^\prime(z) > 0$.
\end{proof}

Given Lemma~\ref{lem: honest-play} we now prove Theorem~\ref{Theorem 1}.

\begin{proof}[Proof of Theorem~\ref{Theorem 1}]
To prove this theorem, it is sufficient to show that $(H, H)$ is a unique pure Nash equilibrium of our game.
To show that, let us consider users $i$, and $j$, and let us recall punishment factors for playing honestly and faulty
\[
P^{t,H}_i = \frac{1 + \sum_{k\in D_i^H} R_k + R_j}{1 +\sum_{k\in D_i} R_k +R_j};
\
P^{t,F}_i = \frac{1 + \sum_{k\in D_i^H} R_k}{1 +\sum_{k\in D_i} R_k +R_j}
\]
As $R_j$ is a positive value; we can see that $P^{t,H}_i>P^{t,F}_i$. Multiplying both sides by $R_i^t$, hence $R_i^t\cdot P_i^{t,H}>R_i^t\cdot P_i^{t,F}$.

Given Lemma~\ref{lem: honest-play}, and the fact that $R_i \in (0,1)$, we can write:
\[
   R_i^t\cdot P_i^{t,H} +f(R_i^t\cdot P_i^{t,H},t_i) 
   >R_i^t\cdot P_i^{t,F} +f(R_i^t\cdot P_i^{t,F},t_i)\nonumber
\]
and similarly for user $j$:
\[
   R_j^t\cdot P_j^{t,H} +f(R_i^t\cdot P_j^{t,H},t_j) 
   >R_j^t\cdot P_j^{t,F} +f(R_j^t\cdot P_j^{t,F},t_j).\nonumber
\]
This is the new reputation score. The outcome of the new reputation score for both users is always greater when a user is contributing honest work. Therefore, playing honestly is a dominant strategy for both users, and $(H, H)$ is a pure Nash equilibrium. This Nash equilibrium is unique because $H$ dominates the only other pure strategy $F$.
\end{proof}

A higher reputation score results in a higher chance of publication in borderline scenarios.  Given Assumption~\ref{assumption: propose}, we can conclude that users are self-incentivized to play honestly and keep a high reputation score.

Now let us consider the $\pi\in[0,1]$ to be the probability that the fault detection oracle detects a fault correctly. Similarly, consider $\bar\pi\in[0,1]$ to be the probability that the honest behavior is detected correctly. With this assumption, let us revisit our game as represented in Table~\ref{table: Peer Review Game}.
We consider $X^\prime$ as the outcome for honest behavior that can be falsely detected as faulty with probability $1-\bar\pi$.  
\[
X^\prime = \bar\pi X + (1-\bar\pi) Y
\]
Similarly, $Y^\prime$ follows but for faulty behavior.
\[
Y^\prime = (1-\pi)X + \pi Y
\]

The following theorem proves that our system rewards honest behavior given an imperfect fault detection oracle.
\begin{myTheorem}
\label{Theorem 2}
Let $\alpha<\frac{1}{6}$ and $\pi,\bar\pi\in [0,1]$ with $\bar\pi>(1-\pi)$.
The outcome $(H,H)$ is the unique pure Nash equilibrium for the game described in Table~\ref{table: Peer Review Game} with an imperfect fault detection oracle.
\end{myTheorem} 
\begin{proof}
From Theorem~\ref{Theorem 1} we know that $X>Y$ and with the assumption $\bar\pi>(1-\pi)$ we get 
\[
  X^\prime = \bar\pi X + (1-\bar\pi) Y >  (1-\pi)X + \pi Y = Y^\prime 
 \]
Hence $X' > Y'$ and $(H,H)$ is the unique pure Nash equilibrium.
\end{proof}
The inequality $\bar\pi>(1-\pi)$ means that the probability of detecting honest behavior is higher than being falsely detected as honest while acting faulty. If around half of the behavior is detected falsely, this inequality still holds.

\subsection{Inclusivity}

Our system ensures inclusivity across venues, giving all papers the chance of getting published based on their quality. In particular, we show that our system provides fairness and recovery properties. 
We will discuss these properties shortly and then show the detailed proofs.

\noindent \textbf{Recovery} A user in \system has the chance to recover from a faulty behavior over time. This is a key element that is incorporated to gain function that we designed. 

\noindent \textbf{Fairness} Our system is designed to give a fair chance to all users. Toward this goal, we benefit from randomization. This
ensures that all users have the chance to be a reviewer, and also a paper would get a diverse set of reviewers. 

\subsection{Recovery}
\label{sec : individual fault}

There is a chance that a user does not behave according to the rules of the system for a period of time, e.g., the reviewer is \emph{lazy}. 
We call a reviewer lazy if that reviewer does a review repeatedly and does not add additional value to what has been done before. Our aim is to avoid such evaluations. 

\noindent \textbf{Lazy review approximation}
\label{sec:Lazy review approximation}
In line with Assumption~\ref{assumption: cheating detection}, we consider that a lazy behavior can be approximately detected by an oracle with a precision probability of $\pi$. The oracle compares the contributions to older ones stored in the blockchain, and a user $i$ submits a lazy review with probability $\mu$. 

We next use a standard assumption that in the blockchain context, the number of such lazy reviewers or other faulty users is below a certain threshold~\cite{ChenM19}.

\begin{myAssumption}
\label{assumption: honest user}
At least $\frac{2}{3}$ of users behave honestly.
\end{myAssumption}

With the above assumption, it is clear that the average user has a reputation score of at least $0.5$ (as the reputation of honest users only increases).  
Assume that a user performs a maximum of $1$ reviewer per time interval for this approximation since a review takes significantly longer than a single time interval. The punishment value, if the user only reviews, is at least  $P^t_i=\frac{1.5-\mu\pi 0.5}{1.5}$.
If we consider $\pi\in[0.5,1)$ and $\mu\in [0.1,1]$, we have $P^t_i \in (0.\overline{666},0.98\overline{3}]$.
An average value for $P^t_i$ in the worst case seems to be very high, but getting punished once has a significant impact as the reputation is decreased by a factor of $\frac{2}{3}$ on average. Recovering from that could take longer if $\alpha$ is chosen to be small.

\noindent \textbf{Recovery from lazy reviews}
A user can only recover in rounds when the user is active. Assume that the user is honest. Continuing the discussion from the previous paragraph, the reputation update on average is: 
\begin{align}
\label{eq: alpha aproximation}
R_i^t>R_i^t\cdot\frac{1.5-\mu\pi 0.5}{1.5}+\alpha
\Leftrightarrow R_i^t>\frac{\alpha}{1-\frac{1.5-\mu\pi 0.5}{1.5}}
\end{align}
 With the calculation of the average punishment and the fact, that $R_i^t\in (0,1)$, one can conclude:
 \[
R_i^t>\frac{\alpha}{1-0,98\overline{3}}=\frac{\alpha}{0.11\overline{6}}
\Rightarrow \alpha<0.11\overline{6}
\]
This approximation (which can change with the chosen parameter) shows that $\alpha$ should be low. Setting $\alpha$ very low extends the domain in which Equation~\ref{eq: alpha aproximation} is true to any reputation score above 0. Setting 
$\alpha < \frac{0.11\overline{6}}{2}$
would mean, that it is not reasonable to do a lazy review when $R_i^t\geq 0.5$. In particular, this means that lazy reviews are costly in this case, as they can lower the reputation score below $0.5$ - the reputation score of a new user. 
This shows that lazy reviews become undesirable with a low $\alpha$ value even in the worst case.

\begin{figure}[t]
    \centering
    \includegraphics[width=0.95\linewidth, clip]{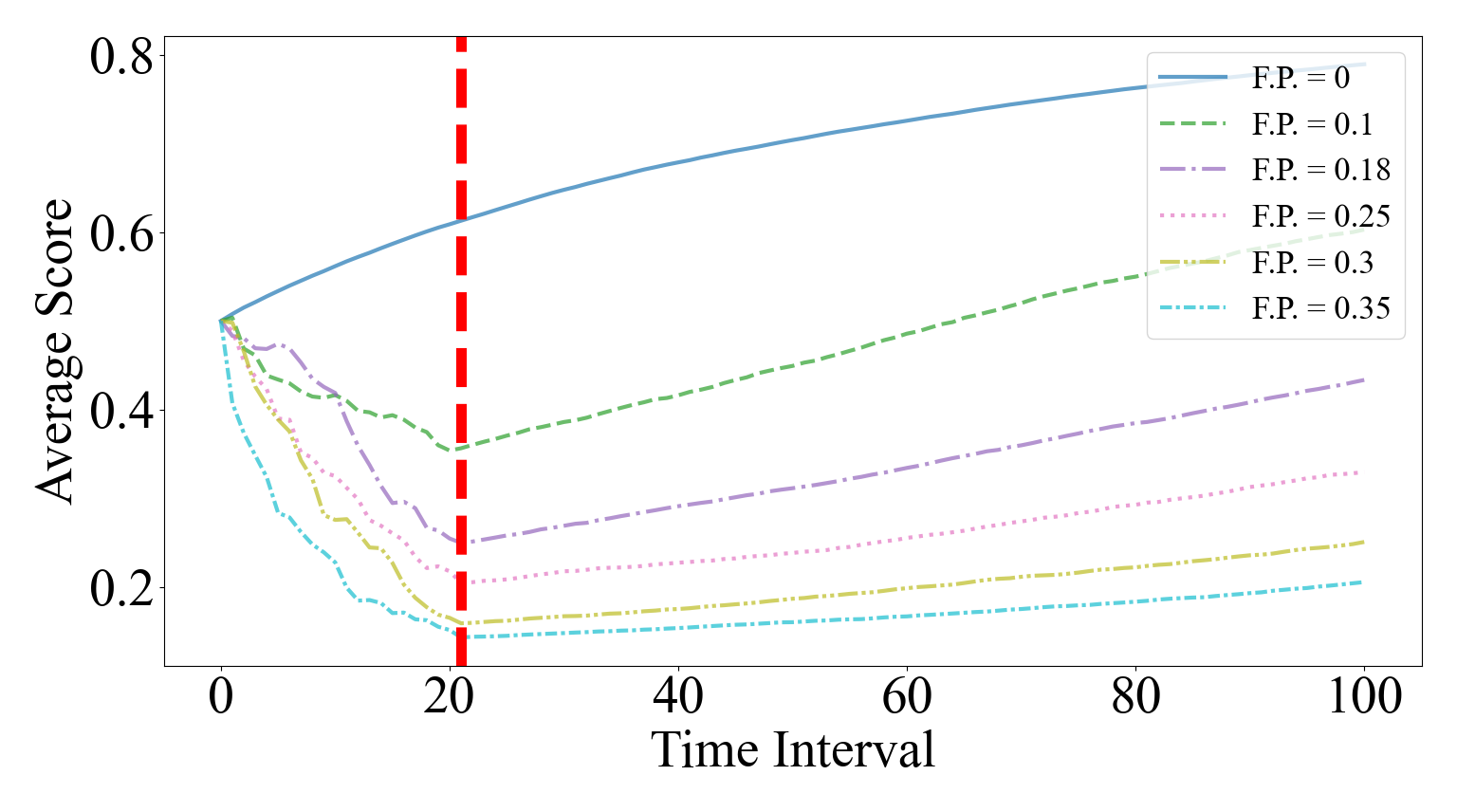}
    \caption{Change in the user score over time, considering the different probability of faulty behavior (F.P.). Each line indicates a different fault probability of a user. 
    }
    \label{fig: recovery}
\end{figure}

In Figure~\ref{fig: recovery}, we visualize an example of how faulty behavior manifests itself for an individual. Each line represents the average reputation of users that are faulty with the same probability. For example, just being faulty $10\%$ of the time has a large influence on the reputation score compared to non-faulty behavior.
In this example, after $20$ time intervals, the behavior is set to be honest for all users, in order to visualize the recovery rate. By system design, it takes more time to gain a reputation than to lose it. This is especially true for low reputation score users. This graph is very sensitive to the value of $\alpha$ since the reward and punishment are directly linked.  

\subsection{Fairness}
Getting the majority of votes in the unified reviewer can be important for a malicious group if the proposals would otherwise not pass the predefined threshold for the average score of the paper. We call such an attempt of trying to get the majority in the unified reviewer called \emph{majority cluster attack}.
As the reviewers are drawn uniformly at random from the pool of reviewers, if the user size is sufficiently large, controlling the unified reviewer and influencing the result is not feasible with high probability. 

Analyzing the success probability of a majority clustering attack allows us to understand the magnitude of its effect.
The probability of a majority clustering attack is calculated based on the number of users, malicious group size, and the required number of users for a majority in the unified reviewer.

 \begin{myLemma}
 \label{lem3: majority cluster attack prob}
 Let $n = |N|$ be the number of users, 
 $g = |G|$ the size of the malicious group among all users, and $m$ the number of users needed to get a majority for the unified reviewer $Tr$. Consider the unified reviewer consists of $r$ reviewers. Then the probability of success of a majority cluster attack is:
\[
P(Tr\cap G \geq m):=\sum_{i=m}^{r} \binom{r}{i}\prod_{k=0}^{i-1}\frac{g-k}{n-k}\prod_{j=i}^{r-1}\frac{n-(g-i)-j}{n-j}
\]
 \end{myLemma}
 \begin{proof}
Consider the probability $P(Tr\cap G =i)$ for a fixed $i\geq m$. There are $\binom{r}{i}$ different combinations to arrange $i$ malicious users in a group of $r$ users. The order of picking a malicious or honest user does not matter in this scenario. Next, consider one possible way to arrange the $i$ users and take one user from the set $N$ after another for the unified reviewer. One can arrange the group in a way that the first $i$ persons are malicious because the order does not matter. In the $k$-th step of picking a member from $G$ there are $g-k$ possibilities when indexing starts from $0$. After picking from the malicious group, there are $n-(g-i)$ users left, and in the $j$-th step of picking from the are $n-(g-i)-j$ possible ways to do so. The probability for this scenario is then: 
\[p_i := \prod_{k=0}^{i-1}\frac{g-k}{n-k}\prod_{j=i}^{r-1}\frac{n-(g-i)-j}{n-j}\]

In the case $i=r$, we have $p_i := \prod_{k=0}^{i-1}\frac{g-k}{n-k}$. Since the order does not matter, the probability of the other arrangement is the same when $i$ is fixed. Therefore, the probability of $i$ malicious user in the unified reviewer is
$P^t_i := \binom{r}{i} p_i$.

By definition $r\leq i\leq m$, hence:
$P(Tr\cap G \geq m) =\sum_{i=m}^{r} P^t_i$.
\end{proof}
In this calculation, it is assumed that the unified author is not contained in the set of users. To adapt the reasoning above, one has to subtract the unified author from the set of users.

\noindent \textbf{Worst case majority clustering attack}
    In the worst case, $\frac{1}{3}$ of the users are conspiring together.
    Assuming that the unified reviewer consists of $5$ users, we have: 
\begin{align}
\label{eq: p_i for proof of lemma}
    p_i = \prod_{k=0}^{i-1}\frac{\frac{1}{3}n-k}{n-k}\prod_{j=i}^{4}\frac{\frac{2}{3}n+i-j}{n-j}
\end{align}

\begin{myTheorem}  
With Assumption~\ref{assumption: honest user} and the unified reviewer consisting of $r=5$ reviewers, the worst-case probability of success for a majority cluster attack converges to
\[
\lim_{n\rightarrow\infty}P(Tr\cap G \geq m) \le \frac{17}{81}
\]
\end{myTheorem}

\begin{figure}[t]
    \centering
\includegraphics[width=0.95\linewidth, clip]{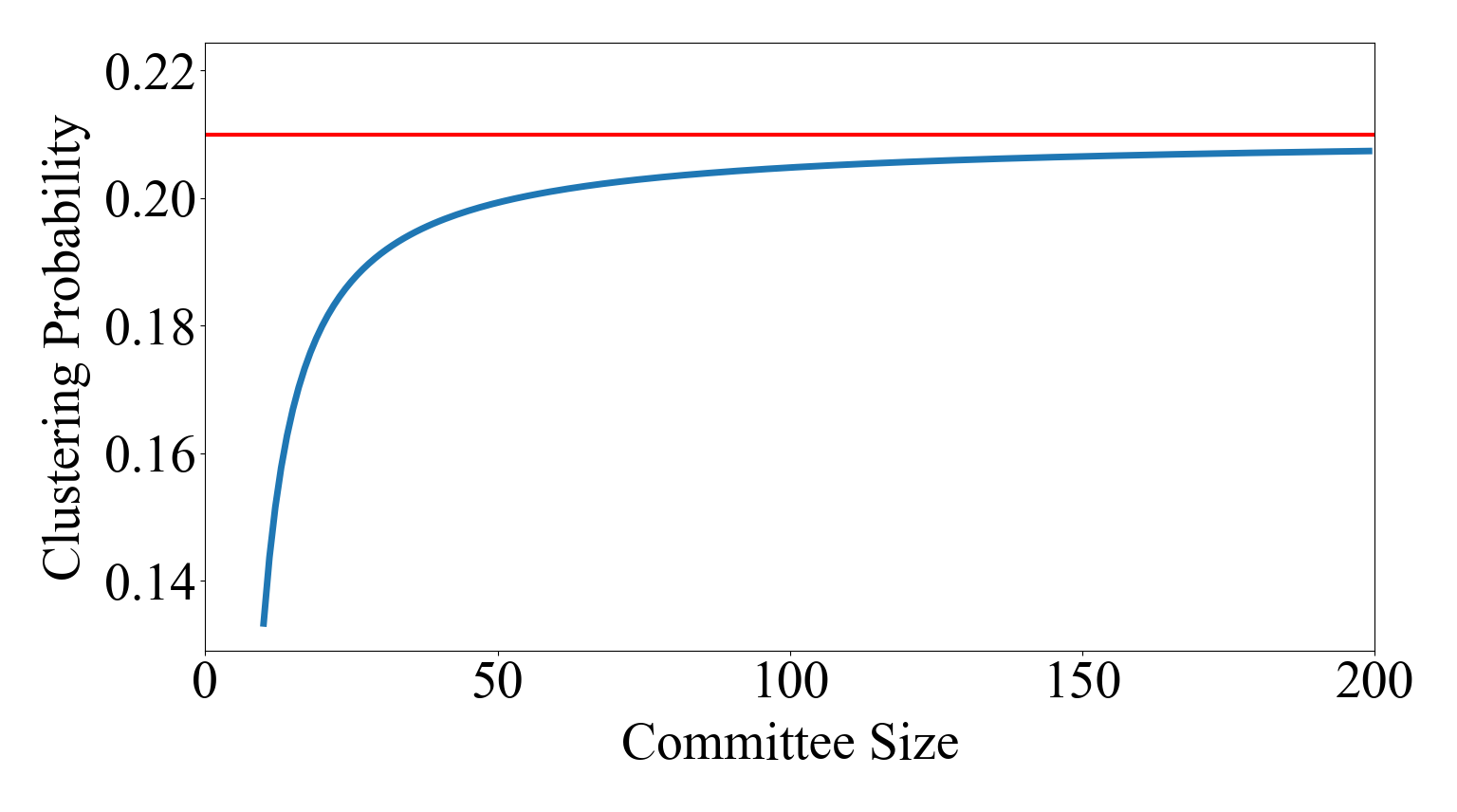}
    \caption{This figure shows clustering probability for a range of committee sizes. The blue line shows how probability evolves, and the red line is the upper bound of probability.} 
    \label{fig: worst case clustering}
\end{figure}

\begin{proof}
Considering Equation~\ref{eq: p_i for proof of lemma}, 
the numerator and denominator are polynomials of degree $5$, both diverge for $n\rightarrow\infty$ and are continuous. Hence, to find the limit, we use the L'Hôpital rule five times as the derivatives are divergent and continuous too. 
Let us reformulate $p_i$ to
\[
 p_i=\prod_{s=0}^4\frac{1}{n-s} \prod_{k=0}^{i-1}\frac{1}{3}n-k\prod_{j=i}^{4}\frac{2}{3}n+i-j
\]
Then we compute it for $i\in\{3,4,5\}$ cases.
For the case $i=3$, we have:
\[LH_3:=\left(\prod_{k=0}^{2}\frac{1}{3}n-k\prod_{j=3}^{4}\frac{2}{3}n+i-j\right)^{\prime\prime\prime\prime\prime}=\frac{160}{81} \]
Similarly for cases $i\in{4,5}$ and for the denominator we have:\\
\[
LH_4=\frac{80}{81}, LH_5=\frac{40}{81}, LH_{den}= 120
\]
With these values, the limit is:
\begin{align*}
    \frac{1}{LH_{den}}\left(\binom{5}{3}LH_3+\binom{5}{4}LH_4+\binom{5}{5}LH_5\right) = \frac{17}{81}
\end{align*}
\end{proof}
Hence, the probability majority clustering attack is low. Figure~\ref{fig: worst case clustering} shows how the worst-case clustering behaves. The probability increases rapidly for $n<50$ and then converges, meaning that we have an upper bound on the probability of success for a majority clustering attack for all $n$.

\subsection{Further Considerations}
Lastly, we detail further considerations that can be taken into account while deploying our proposed peer review system.

\noindent \textbf{Deviation of scores.}
A group of users might decide to score the submissions of each other higher than an honest user does. 
As a response to such a behavior, a random review of reviews of a user over time can be implemented.
If a clear deviation from the average is detected for a certain group of reviewers, then their reviews are considered faulty. Such a review on-reviews method has the benefit that it can distinguish between accidental mishaps and dishonest clusters.

\noindent \textbf{Partial punishment after review comparison.}
A reviewer who does not agree with the average score is the only reviewer who is not lazy; but being punished. To avoid this, we can introduce an overview round: before the reputation is updated, everybody can see the reviews of others and update their version of the review. If the outcome of the review changes, reviewers who previously agreed with average score partially punished. This partial punishment is not as severe as normal dishonest behavior. Let $P^{t,p}_i$ be the partial punishment; therefore, we want to have
$P^{t,p}_i>\frac{(1-\pi)+\pi\cdot P^t_i}{2}$.

Hence, by updating the lazy review, the punishment is less severe on average than the normal punishment.
An exact value for the partial punishment can be defined as:
\[P^{t,p}_i :=  \frac{(1-\pi)+\pi\cdot P^t_i}{2} + \left(1- \frac{(1-\pi)+\pi\cdot P^t_i}{2}\right)\cdot R_i^t\]

\noindent \textbf{Cluster detection.}
An add-on to \system can be cluster detection, in which one aims to detect groups of users collaborating towards a malicious outcome. For a full overview of cluster detection approaches, we refer to~\cite{si2020ReviewOnClusterDetection}. 

\section{Case Study}
\label{sec: case}
We shortly overview our system in the case of a few adversarial review methods.

Firstly, having a reviewer practicing so-called \emph{blind reviewing}~\cite{AdvReview}, that is, reviewing without reading the paper. In such a case, the corresponding review would be phrased in broad terms and would not reflect the content of the paper. Several such reviews coming from the same reviewer are then flagged by a similarity detection mechanism. Therefore, the reputation of this reviewer will be lowered in our system. 
Another example is the \emph{silent but deadly reviewing}~\cite{AdvReview} approach, where the reviewer provides short reviews with minimal explanation for the review score. In that case, such a reviewer can be caught in the review comparison phase and be partially punished.

Other types of adversarial review behaviors can be detected semi-automatically by plugging in additional detection tools to our system. For example by augmenting our system with tools that detect texts created by emerging generative AI systems such as ChatGPT, that are already disrupting educational systems~\cite{lo2023impact}.

In all cases, a faulty action of reviewers lowers their reputation score, which in turn, results in a lower chance for blind reviewers to get their papers accepted in borderline cases, or become part of future reviewing committees.

\section{Conclusion}
\label{sec: conclude}
This paper introduced a self-incentivized, cross-venue, and inclusive peer review system. 
We detailed a mechanism that provably motivates users to follow the rules while submitting and reviewing a paper, ensuring a fair chance for every user to participate in our mechanism. 
Our method relies on a reputation score, which in turn can be utilized across venues, assisting program chairs and the choices they have to make.
We discussed how our system behaves in the presence of faulty behaviors through theoretical reasoning and empirical simulations.
\bibliographystyle{IEEEtran}
\bibliography{peer.bib} 

\end{document}